\documentclass[epj,nopacs]{svjour}
\usepackage{amssymb}
\usepackage{latexsym}
\usepackage{graphics}
\begin{document}
\title{Constraining String Gauge Field by Planet Perihelion Precession and Galaxy Rotation Curves}
%\subtitle{Do you have a subtitle?\\ If so, write it here}
\author{Feng Xu% etc
% \thanks is optional - remove next line if not needed
\thanks{\emph{Present address:} 22 Hankou Road, Nanjing, 210093, China}%
}                     % Do not remove
%
%\offprints{}          % Insert a name or remove this line
%
\institute{Department of Physics, Nanjing University, \email{schyfeng@gmail.com}}
\date{Received: date / Revised version: date}
% The correct dates will be entered by Springer
%
\abstract{
 We performed various tests on a cosmological model in which the string gauge field, and its
coupling to matter, is used to explain the rotation dynamics of stars in a galaxy. Observations used
include perihelion precession and galaxy rotation curves. We solved precession motions for
perturbations of 1) a Lorenz-like force and 2) a general power law central force. For the latter
case, a simple rule judging pro- or retrograde motions is derived. We attributed the precession of
solar system planets to the force due to the string field and calculated the field strengths. We
then fitted the resultant field strengths with a profile consisting one part generated by the Sun
and another background due to other matter in the Milky Way. We used the Milky Way rotation curve to
estimate the field strength and compared it with the one found by precession. The field strengths in
another 22 galaxies are also analyzed.  The strengths form a range that spans 2 orders of magnitude
and contains the value found for the Milky Way. We also deduced from the model a relation between
field strength, galaxy size and luminosity and verified it with data of the 22 galaxies.
\PACS{
      {PACS-key}{discribing text of that key}   \and
      {PACS-key}{discribing text of that key}
     } % end of PACS codes
} %end of abstract
\maketitle
\section{Introduction}
\label{sec:introduction}
This is a paper analyzing a model proposed in \cite{CheungGRC0:2007}. The model is used
as an alternative of dark matter to solve the galaxy rotation curve problem. It
is conjectured that the gauge field in string theory provided the extra 
force needed to keep the rotation curves flat. Just like an electrically charge
particle in magnetic field, the stringly charged matter in the string gauge
field would also feel a force.
In \cite{CheungGRC1:2008}, the string model was put into a contest with a particular dark 
matter model by fitting 22 galaxy rotation curves. These two models showed
similiar fitting power. In this paper we further
test this model with planet precessions in the solar system. In \cite{Iorio:2009}, possible
anomalous
precession for the Saturn was reported, and no explanation within convensional 
description of the solar system seems exist. It is thus
interesting to ask if the effect of the string gauge field could explain the remaining anomalous
precession. That serves as
a test of the string model independent from that with rotation curves. For the consistency of the
string model, it is also important to compare the field strength got by different methods and
observations. 
\paragraph{}
In section \ref{sec:precessions} we prepare the mathematics for precession analysis. 
Using Laplace-Runge-Lenz vector \cite{Goldstein:mechanics}, we derive the precession rate formula 
for a magnetic like force. In addition, the
case of power law central force is also analyzed and we find a simple precession
formula, which is then tested in the Newtonian limit of Schwarzschild space-time
and compared with other analysis on dark matter for the solar system.
\paragraph{}
In section \ref{sec:precession_solar_system} we attribute unexplained anomalous precessions to the
magnetic like
force in the string model. Using formulas derived, we use the precession data to
find the string field strength in the solar system. The field strength is found
to be on the order of $10^{-17}s^{-1}$ (with appropriate dimension, see below
for the detail). Except at the Saturn, a decreasing pattern of the field 
strength with distances to the Sun is found. Other effects of the string field 
with that strength are also discussed. 
\paragraph{}
The decreasing pattern of the field is further explored in the section 
\ref{sec:profile_fitting}. Assuming the string field in the solar system 
consists of one power law part from the Sun and a constant background
from other matter in the Milky Way, a profile fitting of the field strength is
done. The part from the Sun is found to go as $r^{-3}$, like a magnetic dipole.
The background is found to be of opposite sign of the part from the Sun. This
correctly matches with the fact that the rotation direction of the solar system
is opposite to the one of the Milky Way. 
\paragraph{}
To compare with the result from fitting precession, in section
\ref{sec:GRC_milkyway} we use the
rotation curve of the Milky Way to estimate the field strength in the Milky
Way. The connection of this result with the one from precession is then discussed.
\paragraph{}
In section \ref{sec:22galaxy} we first discuss the fitting result in \cite{CheungGRC1:2008} for
field 
strengths of 22 galaxies, emphasizing connection with the current paper. The 
field strength was found to be in the range\cite{CheungGRC1:2008}
$6\times10^{-18}s^{-1}\sim 1\times10^{-15}s^{-1}$. Based on an analogy with electromagnetism, for
the string model we continue to 
derive a relation between $\Omega\cdot R^2_0$ and $L$, where $\Omega$ is the 
string field strength (in appropriate dimension), $R_0$ and $L$ are the size 
scale and the luminosity of the galaxy, respectively. This relation is then 
tested and verified by the data from \cite{CheungGRC1:2008}.

\section{Precessions}
\label{sec:precessions}
The basic tool used for the precession analysis is the Laplace-Runge-Lenz vector defined
by\cite{Goldstein:mechanics,Weinberg:gravitation}
\begin{eqnarray}
  \label{eq:Runge-Lenz_vector}
  \vec{A}=\vec{v}\times \vec{h}-GM_{\odot}\frac{\vec{x}}{r}
\end{eqnarray}
where $\vec{h}$ is the orbital angular momentum per mass of the planet under
consideration, $\vec{x}$ and $r$ is respectively the position vector from the
sun and its magnitude. If the perturbation force is small enough such that the
orbit becomes a precessing ellipse, we have\cite{Weinberg:gravitation}
\begin{eqnarray}
  \label{eq:time_derivative_of_Runge-Lenz_vector}
  \frac{d \vec{A}}{d t}&=& \vec{\eta}\times \vec{h}+\vec{v}\times
  (\vec{x}\times \vec{\eta})
\end{eqnarray}
where $\vec{\eta}$ is the extra \emph{acceleration} due to the perturbing
force.
\paragraph{}
 To compute the angle precessed in one period, we need to
integrate this quantity over time of one period and the angle precessed is then 
$\left|\frac{\Delta \vec{A}}{|\vec{A}|}\right|$. The precession direction can be
read from the direction of $\Delta\vec{A}$. Note that from
(\ref{eq:time_derivative_of_Runge-Lenz_vector}) the time derivative of the
Laplace-Runge-Lenz vector is linear in the perturbation acceleration, thus we can
consider precessions from different perturbation forces seperately.
Here it must be stressed that all
orbital quantities such as $\vec{x}$, $\vec{v}$ and $\vec{h}$ used to do the
integral are that of the \emph{unperturbed} perfect elliptic orbit. Conceptually
we are pretending that the perturbation do not affect the orbit at all in one
period. The integration collects and stores the effect and we then put back the
total effect at one shot when the planet returns to its perihelion. 
\begin{figure}
\begin{center}
% Use the relevant command for your figure-insertion program
% to insert the figure file.
% For example, with the option graphics use
\resizebox{0.4\textwidth}{!}{%
  \includegraphics{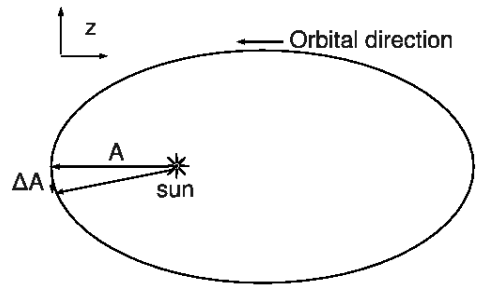}
}
% If not, use
%\vspace{5cm}       % Give the correct figure height in cm
\caption{Laplace-Runge-Lenz vector, its variation and the orbital plane as
 the complex plane}
\label{fig:complex_plane}       % Give a unique label
\end{center}
\end{figure}

\paragraph{}
Since the force due to the string gauge field is magnetic like \cite{CheungGRC0:2007}, we
first check the precession caused by magnetic like force. After that we
continue to analyze the effect of general power law central force.

\subsection{Magnetic like force}
\label{subsec:magnetic_like_force}
For magnetic like force we have
\begin{eqnarray}
  \label{eq:extra_acceleration}
  \vec{\eta}&=& \frac{q}{m}\vec{v}\times \vec{B}
\end{eqnarray}
Assuming the magnetic field is pointing upward perpendicular to the solar
plane in which all planets move counterclockwise\footnote{All planets in
the solar system move counterclockwise if we watch their motion from the side of
the solar plane where we can see the northern hemisphere of the earth.}, we have
\begin{eqnarray}
  \label{eq:time_derivative_Runge-Lenz_vector_magnetic_field}
  \frac{d \vec{A}}{dt}&=& -\frac{q}{m}Bh \vec{v}-\frac{q}{m}(\vec{x}\cdot
  \vec{v})(\vec{v}\times \vec{B})
\end{eqnarray}
and for integral over time, denoted by $\Delta \vec{A}$,
\begin{eqnarray}
  \label{eq:integral_Rung-Lenz_vector_magnetic_field}
  \int d \vec{A}= -\frac{q}{m}\int Bh \vec{v}dt-\frac{q}{m}\int
  (\vec{x}\cdot \vec{v})(\vec{v}\times \vec{B})dt
\end{eqnarray}
Firstly let us check the first term. Because planet orbits all have very small 
eccentricities, we can assume a constant magnetic field
over the whole orbit for \emph{each particular} planet. Thus $B$ can be taken 
outside of the integral. The factor $h$ can also be taken outside because it 
depends only on the orbit and it is constant for the unperturbed orbit 
\footnote{All orbital quantities are those 
for the unperturbed orbit. }.
Therefore the first term is proportional to 
\begin{eqnarray}
  \label{eq:first_term_of_integral}
  \int \vec{v}dt =\int d \vec{x}=\Delta \vec{x}
\end{eqnarray}
and vanishes for unperturbed orbit. Now we are left with only the second
term. It can be simplified as follows 
\begin{eqnarray}
  \label{eq:2ed_term_of_integral}
  &&\int (\vec{x}\cdot \vec{v})(\vec{v}\times \vec{B})dt\\\nonumber
  &=& \int (\vec{v}\times
  \vec{B})d\left( \frac{1}{2}\vec{x}^2 \right)
  = -\int \frac{1}{2}\vec{x}^2 d(\vec{v}\times \vec{B})\\\nonumber
  &=& \vec{B}\times \int \frac{1}{2}\vec{x}^2 \frac{d \vec{v}}{dt}dt
  = \vec{B}\times \left( -\frac{k}{2}\int \hat{x}dt \right)\\
  &=& \vec{B}\times \left( -\frac{k}{2}\int \hat{x}\frac{1}{\omega} d\theta
  \right)=\vec{B}\times \left( -\frac{k}{2h}\int \hat{x}r^2 d\theta
  \right)
\end{eqnarray}
where we have used partial integration and Newton's second law
$\vec{a}=-\frac{k\hat{x}}{|\vec{x}|^2}$ and $\omega$ is the angular velocity.
Regarding the planetary motion plane as a complex plane, we can denote
$\hat{x}$ by $\exp(i\theta)$, and for elliptic orbit with semimajor axis
$a$ and eccentricity $e$, $r=\frac{a(1-e^2)}{1-e\cos \theta}$, therefore
\begin{eqnarray}
  \label{eq:simplified_2ed_term_of_integral}
  \int \hat{x} r^2 d\theta = \left[ a(1-e^2) \right]^2 \int
  \frac{\exp(i\theta)d\theta}{(1-e\cos\theta)^2}
\end{eqnarray}
Combining above formulas, we get
\begin{eqnarray}
  \label{eq:variation_of_A_1circle_not_integrated}
  \Delta \vec{A}&=& i \frac{qkB}{2hm}\left[ a(1-e^2) \right]^2 f(e)
\end{eqnarray}
where 
\begin{eqnarray}
  \label{eq:f_integralform}
  f(e)&=& \int_0^{2\pi}\frac{\exp(i\theta)d\theta}{(1-e\cos\theta)^2}
\end{eqnarray}
Remembering that $\Delta\phi=\left| \frac{\Delta\vec{A}}{|\vec{A}|}\right|$,
and \cite{Goldstein:mechanics,Weinberg:gravitation}
\begin{eqnarray}
  \label{eq:various_quantities_for_delta_phi}
  k&=& GM_{\odot}\\
  L&=& hm=\sqrt{GM_{\odot}m^2a(1-e^2)}\\
  |\vec{A}|&=& eGM_{\odot}
\end{eqnarray}
we get
\begin{eqnarray}
  \label{eq:deltaPhi1}
  \Delta\phi=-\left( \frac{qB}{m} \right)\frac{1}{\sqrt{GM_{\odot}}}
  \left[ a(1-e^2) \right]^{\frac{3}{2}}\frac{f(e)}{2e}
\end{eqnarray}
$f(e)$ can be integrated exactly and is
\begin{eqnarray}
  \label{eq:fe}
  f(e)=2\pi e(1-e^2)^{-\frac{3}{2}}
\end{eqnarray}
and thus
\begin{eqnarray}
  \label{eq:deltaPhi_final}
  \Delta\phi=-\left( \frac{qB}{m} \right)\frac{1}{\sqrt{GM_{\odot}}}\pi
  a^{\frac{3}{2}}
\end{eqnarray}
This is the angle precessed in one revolution, the precession rate in time is 
\begin{eqnarray}
  \label{eq:precession_rate}
  \delta \dot{\omega}=-\left( \frac{qB}{m} \right)\frac{1}{\sqrt{GM_{\odot}}}\pi
  \left( \frac{a^{\frac{3}{2}}}{T} \right)
\end{eqnarray}
Note that the eccentricity disappeared here. Also note that one of Kepler's law
says that $\frac{a^{\frac{3}{2}}}{T}$ is a constant for planets in the solar
system. But in later calculation we will not treat it as a constant but simply 
use the data of $a$ and $T$ to compute it.

\subsection{Power law central force}
\label{subsec:power_law_central_force}
For central force the time derivative of Laplace-Runge-Lenz vector simplies to 
\begin{eqnarray}
  \label{eq:Runge_Lenz_time_derivative_simplied_for_central_force}
  \frac{d \vec{A}}{dt}=\vec{\eta}\times \vec{h}
\end{eqnarray}
and thus
\begin{eqnarray}
  \label{eq:Runge_Lenz_variation_in_one_revolution}
  \Delta \vec{A}&=& \int \vec{\eta}\times \vec{h} dt=h\int
  \eta(r)\hat{x}\times \hat{z} dt\\
  &=& \int \eta(r)r^2\left( \hat{x}\times \hat{z} \right)d\theta
\end{eqnarray}
where $\eta$ is the magnitude of the perturbation acceleration, outward from the
center is defined to be positive. For general power law central force
$\eta(r)=Kr^n$, using elliptic orbit equation
$r=\frac{a(1-e^2)}{1-e\cos\theta}$ and after some similar algebra as above, we 
get
\begin{eqnarray}
  \label{eq:Runge_Lenz_variation_power_law}
  \Delta \vec{A}=(-i)K\left[ a(1-e^2) \right]^{n+2}f_n(e)
\end{eqnarray}
where, as before, we used a complex number to represent the variation, and 
$f_n(e)$ is a function of eccentricity only, defined by
\begin{eqnarray}
  \label{eq:f_n(e)_definition}
  f_n(e)=\int^{2\pi}_0 \frac{\cos\theta d\theta}{\left( 1-e\cos\theta
  \right)^{n+2}}
\end{eqnarray}
This integral can be done exactly, the key equation is 
\begin{eqnarray}
\nonumber&&\int^{2\pi}_0
\frac{e^{i\theta}d\theta}{\left( 1-\lambda\cos{\theta}
\right)^n}\\
&=&\frac{2\pi}{(n-1)!}\left( -\frac{2}{\lambda} \right)^n
\left( \frac{d}{dz} \right)^{n-1}
\left( \frac{z}{z-z_+} \right)^n |_{z=z_-}
\end{eqnarray}
where $z_\pm=\frac{1}{\lambda}\pm \sqrt{\frac{1}{\lambda^2}-1}$.
But an expasion in $e$ will be 
more illuminating. For planet orbits, we have $e\ll1$, and we can expand the
factor $\frac{1}{(1-e \cos\theta)^{n+2}}$ to 
\begin{eqnarray}
  \label{eq:integral_denominator_expansion}
  1-(n+2)(-e\cos\theta)+\mathcal{O}(e^2)
\end{eqnarray}
Then to the first order of $e$,
\begin{eqnarray}
  \label{eq:f_n(e)_firstorderof_e}
  f_n(e)\approx \int^{2\pi}_0 (n+2)e\cos^2\theta d\theta=(n+2)e\pi
\end{eqnarray}
and for the Laplace-Runge-Lenz vector variation we have
\begin{eqnarray}
  \label{eq:Runge_Lenz_variation_power_law_firstorder_value}
  \Delta \vec{A}\approx (-i)K\left[ a(1-e^2) \right]^{n+2}(n+2)e\pi
\end{eqnarray}
Recalling $|\vec{A}|=eGM_\odot$, the angle precessed in one revolution
$\Delta\phi=|\frac{\Delta \vec{A}}{\vec{A}}|$ is
\begin{eqnarray}
  \label{eq:precession_angle_one_revolution_power_law_central_force}
  \Delta \phi\approx \frac{K\pi}{GM_\odot}\left[ a(1-e^2)
  \right]^{n+2}(n+2)
\end{eqnarray}
This precession angle formula is for the most general power law central 
force perturbations. Following are some example applications.
\paragraph{Schwarzschild space-time}
For Schwarzschild space-time a test particle's orbit obeys\cite{Misner:gravitation}
\begin{eqnarray}
  \label{eq:EOM_Schwarzschild_metric}
  \tilde{E}^2-\left( 1-2 \frac{M}{r} \right)\left( 1+
  \frac{\tilde{L}^2}{r^2} \right)=\dot{r}^2
\end{eqnarray}
where $\tilde{E}$ and $\tilde{L}$ are the energy and the angular momentum
per unit mass, respectively. After expanding it is
\begin{eqnarray}
  \label{eq:EOM_Schwarzschild_expanded}
  \left( \tilde{E}^2-1
  \right)+\frac{2M}{r}-\frac{\tilde{L}^2}{r^2}+\frac{2M
  \tilde{L}^2}{r^3}=\dot{r}^2
\end{eqnarray}
On the other hand, in classical mechanics we have
\begin{eqnarray}
  \label{eq:EOM_classical_central_force}
  \frac{1}{2}m\dot{r}^2+V+\frac{L^2}{2mr^2}=E
\end{eqnarray}
or
\begin{eqnarray}
  \label{eq:EOM_classical_central_force_expanded}
  \frac{2E}{m}-\frac{2V}{m}-\frac{L^2}{m^2 r^2}=\dot{r}^2
\end{eqnarray}
Comparing this to eq.(\ref{eq:EOM_Schwarzschild_expanded}), it can be read out that
in the Newtonian limit, we should have
\begin{eqnarray}
  \label{eq:classical_Schwarzschild_correspondence}
  \tilde{E}^2-1&=&  \frac{2E}{m},\\
  \frac{\tilde{L}^2}{r^2}
  &=& \frac{L^2}{m^2r^2}\\  
  \frac{2M}{r}+\frac{2M \tilde{L}^2}{r^3}&=& -\frac{2V}{m}
\end{eqnarray}
from the last of the three equations we know, with some dimensional
conversion, the total gravitational potential in the Newtonian limit is
\begin{eqnarray}
  \label{eq:effective_potential_Schwarzschild_total}
  \Phi=-\frac{GM}{r}-\left( \frac{GM}{c}
  \right)^2a(1-e^2)\frac{1}{r^3}
\end{eqnarray}
The first term is the usual Newtonian potential. The second term considered as 
the perturbation potential causing precession gives acceleration
\begin{eqnarray}
  \label{eq:perturbation_acceleration_Schwarzschild}
  \eta=-\left( \frac{GM}{c} \right)^2a(1-e^2)\frac{3}{r^4}
\end{eqnarray}
thus $K=-3\left( \frac{GM}{c} \right)^2a(1-e^2)$ and $n=-4$. By
(\ref{eq:precession_angle_one_revolution_power_law_central_force})
the corresponding precession per revolution is 
\begin{eqnarray}
  \label{eq:precession_Schwarzschild}
  \Delta\phi= \frac{6\pi GM_\odot}{c^2}\frac{1}{a(1-e^2)}
\end{eqnarray}
This is precisely the one directly derived from general 
relativity\cite{Misner:gravitation}, e.g it is the $43''/cy$ for the planet Mercury.
\paragraph{Solar system with dark matter}
Equation (\ref{eq:precession_angle_one_revolution_power_law_central_force}) can
also be used to derive precession caused by possible dark matter spherically
distributed in the solar system. In this case we have 
\begin{eqnarray}
  \label{eq:extra_acceleration_dark_matter_solar_system}
  \eta(r)=-\frac{4\pi}{3}\rho_0Gr
\end{eqnarray}
where $\rho_0$ is the dark matter density, and the corresponding precession is 
\begin{eqnarray}
  \label{eq:precession_dark_matter_solar_system}
  \Delta\phi=-\frac{4\pi^2 \rho_0}{M_\odot}\left[ a(1-e^2) \right]^3
\end{eqnarray}
After some dimensional conversion, this result is the same with the one 
in \cite{Gron:1995} except that the factor $(1-e^2)$ is absent there. It is
because \cite{Gron:1995} considered only circular orbit. The result here
works also for ellipse. 
\paragraph{}
Generally the equation 
(\ref{eq:precession_angle_one_revolution_power_law_central_force})
provides a convenient way to judge whether a power law perturbation causes 
prograde or retrograde precession. $n=-2$ is a critical value. The result is 
summarized in table \ref{tab:retro_or_pro_summary}.
\begin{table}
\begin{center}
  \begin{tabular}{c||c|c|c}
    & $n<-2$ & $n=-2$ & $n>-2$\\\hline
    $K<0$ & $+$& $0$& $-$\\
    $K>0$ & $-$ & $0$ & $+$
  \end{tabular}
  \caption{\label{tab:retro_or_pro_summary}Sign of $\Delta\phi$ for perturbations of form
  $\eta(r)=Kr^{n}$. ``$+$'' for prograde, ``$-$'' for retrograde.}
\end{center}
\end{table}

\section{String field strength from precession in solar
system}
\label{sec:precession_solar_system}
\begin{figure}
\begin{center}
% Use the relevant command for your figure-insertion program
% to insert the figure file.
% For example, with the option graphics use
\resizebox{0.4\textwidth}{!}{%
  \includegraphics{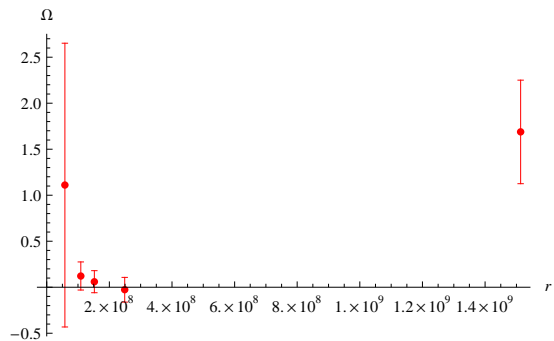}
}
% If not, use
%\vspace{5cm}       % Give the correct figure height in cm
\caption{$\Omega(1\times 10^{-17}s^{-1})$ versus distance
  $r(km)$ to the Sun}
\label{fig:omega_r}       % Give a unique label
\end{center}
\end{figure}

Here we will use the anomalous precession data to calculate the string field 
strength in the solar system. We attribute all the anomalous precession
\cite{Iorio:2009} to the magnetic like force due to the string field. 
We are interested in the
quantity\cite{CheungGRC0:2007}
\begin{eqnarray}
  \Omega=\frac{QH}{m}
  \label{eq:string_omega_definition}
\end{eqnarray}
where $\frac{Q}{m}$ and $H$ are the string charge-to-mass ratio and field 
strength, respectively. $\Omega$ is the string field strength in dimension 
$s^{-1}$. The corresponding one in electromagnetism is $\frac{eB}{m}$. In
this dimension, we can compared it with strengthes of other magnetic-like
forces. We just need to convert the strength into the dimension $s^{-1}$, e.g
$\frac{eB}{m}$ for electromagnetism.
To calculate the strength from precessions we invert (\ref{eq:precession_rate}) and
get
\begin{eqnarray}
  \Omega=(-\delta \dot{\omega})\sqrt{GM_{\odot}}\frac{T}{\pi
  a^{\frac{3}{2}}}
  \label{eq:Omega_for_fitting}
\end{eqnarray}
Because relative errors from other parts are all tiny compared to the one of the
precession rate, the error of $\Omega$ can be computed by
\begin{eqnarray}
  Err(\Omega)=Err(\delta \dot{\omega})\sqrt{GM_{\odot}}\frac{T}{\pi
  a^{\frac{3}{2}}}
  \label{eq:error_omega}
\end{eqnarray}
or simply
\begin{eqnarray}
  \left| \frac{Err(\Omega)}{\Omega}\right |=
  \left |\frac{Err(\delta\dot{\omega})}{\delta\dot{\omega}}\right|
  \label{eq:relative_error_equal_omega_and_Omega}
\end{eqnarray}
The value of $\Omega$ calculated are shown in table \ref{tab:Omega}. And the
diagram is shown in Fig.\ref{fig:omega_r}. Orbital datas are from \cite{HORIZON}. Precession datas
are from \cite{Iorio:2009}.

\paragraph{}
The center values of $\Omega$ for the inner planets exhibit a decreasing
pattern with respect to $r$, although long error bars allow also the case of
vanishing string field\footnote{The same argument applies to precession rate as
well, which is proprotinal to $\Omega$.}. At the Saturn, $\Omega$ is nonzero
within one $\sigma$. However, as mentioned in \cite{Iorio:2009}, the
error bar at the Saturn may actually be bigger, in which case the value of
precession or $\Omega$ then vanish. More precise experiments on precessions are
needed to get more definite conclusions regarding $\Omega$ (or anomalous
precessions) in the solar system. Inspired by this decreasing pattern, in the
next section we will try to fit data of inner planets with a (nearly) power law
profile.
\paragraph{}No matter how critically we take the calculated result of $\Omega$ 
here, it is certainly true that the upper limit of $\Omega$ 
at the solar system is on the order 
of $10^{-17}s^{-1}$. As a compare, let us note that for the real magnetic field
near earth, since $\frac{e}{m}=1.76\times 10^{11}C/kg$ and the
interplanetary magnetic field there is on the order of
\footnote{See the appendix for the details.} 
$10^{-9}T$, we have $\Omega_{em}\sim 100s^{-1}$. In that sense, the string field
strength is $10^{19}$ times smaller than the real magnetic field. 
Then arises the question that  
why we not use the real magnetic field directly to explain the precession, now
that we find the required field strength is so much smaller than observed real
magnetic field. One reason is that matter is electrically neutral,
but assumed in our model to be stringly charged. The real magnetic field can act
on neutral matter only through dipole dipole interaction, which, as explained 
in the appendix, for several reasons does not contribute to planet precession. 
Another question about this string model is, now we are assuming matter
is stringly charged and being acted on by the corresponding magnetic field for
this charge, then is there or where is an electrical interaction 
between stirngly charged matter. After all, so far we seem to be assuming all
matter take the ``same'' (if there are more than one kind of charge) kind of 
charge. This question lies outside of our current model and need more
theoretical investigations into this string charge. As for the model used here,
we \emph{can} say that we are using an electromagnetic-like model with magnetic 
field interactions only. 
\paragraph{}
It is interesting to calculate the
extra acceleration due to the string field for objects on the earth. Since the
field strength is got for the rest frame relative to the Sun, the velocity of
objects on the earth should be almost the same with the velocity of the 
earth relative to the Sun, i.e $30km/s$. The corresponding acceleration produced
is $\sim 10^{-13}m\cdot s^{-2}$. 
\begin{table*}
\begin{center}
  \begin{tabular}{l||c|c|c|c|c}
    \hline
    Name & Mercury & Venus & Earth & Mars & Saturn\\ \hline
    $\delta\dot{\omega}(10^{-4}~''/cy)$ & $-36\pm50$ & $-4\pm5$ & $-2\pm4$ & 
    $1\pm5$ & $-60\pm20$\\ \hline
    T(y) & 0.240846 & 0.6151970 & 1.0000175 & 1.8808 & 29.4571\\ \hline
    a(km) & 57909100 & 108942109 & 152097071 & 249209300 & 1513325783\\ \hline
    e & 0.205630 & 0.0068 & 0.016710219 & 0.093315 & 0.055723219 \\ \hline
    $\Omega(10^{-17}s^{-1})$ & $1.11$ & $1.22\times10^{-1}$ &
    $5.99\times10^{-2}$ & $-2.69\times10^{-2}$ & $1.68794$\\
    \hline
  \end{tabular}
  \caption{\label{tab:Omega}Planets data and values of $\Omega$}
\end{center}
\end{table*}
\paragraph{}
More on the Saturn.
One thing special about the Saturn is that it 
belongs to gas giant while all other planets
considered here are small and solid planets of the inner planet family of the
solar system. As in electromagnetic theory, the content and structures of
planets may affect their interactions with the string field, which
might explains  
the anomalous behavior of the Saturn. This argument can be supported if
anomalous precession behaviors similar to the Saturn's can be observed on other outer planets.
% BibTeX users please use
% \bibliographystyle{}
% \bibliography{}

\section{Profile fitting of precession in solar
system}
\label{sec:profile_fitting}
The decreasing pattern of $\Omega$ for inner planets with respect to $r$ 
indicates it might be useful to fit these values with a power law term.
Presumably we could attribute this $r$ dependent term to the Sun from which
$r$ is measured. On the other hand, note that $\Omega$ at
the Mars is negative, although it seems also sitting on the same curve passing 
the
first three inner planets. One possible configuration for $\Omega$ then is that, in 
addition to the power law term, there is also a weak constant
background $\Omega$ with opposite direction to the $\Omega$ from the
Sun. The background might be provided by all other matter in the universe. The Milky
Way should be the most important source of influence.
Thus here we try to fit field strength at different inner planets with the following 
profile,
\begin{eqnarray}
  \Omega(r)=A+B r^{-a}
  \label{eq:Omega_profile}
\end{eqnarray}
For the actual fitting on computer, the profile used is 
\begin{eqnarray}
  \Omega(r)=\Omega_0+\Omega_1 \left( \frac{r}{10^7 km} \right)^{-a} 
  \label{eq:Omega_profile_actual}
\end{eqnarray}
The best fitting parameters for this profile is 
\begin{eqnarray}
  \Omega_0&=& -0.0223787\times 10^{-17}s^{-1}\\\nonumber
  \Omega_1&=& 260.504\times 10^{-17}s^{-1}\\\nonumber
  a&=& 3.09953
  \label{eq:best_profile_fitting_parameters}
\end{eqnarray}
It is interesting to know the relative strength of the two components from the 
Sun and the constant background, the result is shown in table
\ref{tab:relative_strength_table}. 
\begin{table}
\begin{center}
  \begin{tabular}{l||c|c|c|c}
    \hline
    Planet& Mercury & Venus & Earth & Mars \\
    \hline
    ratio & 50.3 & 7.1 & 2.5 & 0.54 \\
    \hline
  \end{tabular}
  \caption{\label{tab:relative_strength_table}relative strength: power law term/constant term}  
\end{center}
\end{table}
The power law term decreases with $r$ quickly relative to the background. 
We can say that in most areas in the solar system, the string field would just 
be around that background, which is on the order of $10^{-19}$Hz. 
\paragraph{}
Note that $a$ is found to be near 3, which is exactly the
power for a dipole field. It means that the string field interaction between the sun and planets is
similar to that between a magnetic dipole and charged particles.
\paragraph{}
Note that the fitting tells us the constant background in the solar
system is negative. This is good news for the string model. As in
electromagnetism, we expect the string gauge field in a galaxy be generated
by the rotation of (stringly charged) matter in the galaxy, just like rotating
electric charge would generate magnetic field. Because the Sun and
planets in the solar system rotate in opposite direction of that of stars' 
rotation in the Milky Way, therefore naively we would expect the background 
field to be negative
if we consider the one from the Sun as positive. And this is exactly what the
profile fitting told us. Only data in the solar system was used, but 
the conclusion is for the whole Milky Way, specifically for its rotation direction. However,
this conclusion is not so robust as it might seem. Firstly, the long error bars of precession weaken
the conclusion from profile fitting. Secondly, since the Sun contains 99\% of the total
mass in the solar system, all planets are well outside of the major mass
concentration of the solar system, but we do not have enough experimental data
to say whether or not the solar system lies in the major mass concetration of
the Milky Way. As in ordinary electromagnetic theory, for a right handed 
current disk the magnetic field is downward outside of the major current
distribution, but upward if we go into the current disk somewhere, specifically
at the center of the disk. There is a place within the concentration where the
magnetic field changes its sign.\footnote{Considering this, the $\Omega$ in
\cite{CheungGRC0:2007,CheungGRC1:2008} are position-averaged one over the galaxy.}
Thus even if the Milky Way is rotating in opposite
direction from the solar system, if the Sun is too close inside the major mass
concentration of the Milky Way, the background therefrom should still be
positive. If we believe the negative background from the profile fitting, then it necessarily
 implies that the solar system
does not sit very close to the center of the Milky Way. This is reasonalble
assumption but can not be verified for lack of precise data for mass 
distribution in the Milky Way.

\section{$\Omega$ from rotation curve of the Milky way}
\label{sec:GRC_milkyway}
Above from anomalous precession we have obtained $\Omega$ in the solar system. 
The background value there in large part should come from other matter in the 
Milky Way altogether. On the other hand, by the same idea used in
\cite{CheungGRC1:2008}, we can also estimate $\Omega$ at the solar system by the rotation
curve of the Milky Way. A natural check of the string field model would be to
compare $\Omega$ from precessions in the solar system with $\Omega$ from
rotation curve of the Milky Way.
\paragraph{}As mentioned above, we do not have sufficient data to know if the 
solar system is far away enough from the Milky Way center such that its velocity
is increased by the string field. If we trust the negative background got
from precession fitting, the velocity of the solar system in the Milky Way is 
increased by the string field.
(It means the sun is in the lifted part of the rotation curve.) In
the rest part of this section we will assume the velocity of
the solar system is increased by the string field.
\paragraph{}Using the idea in \cite{CheungGRC1:2008}, we can make a rough estimate of 
$\Omega$ in the Milky way as follows. In the string model, the total force on
the sun is composed of only the gravitational attraction from visible mass in
the Milky Way and the magnetic-like force from the string field. The
gravitational force decreases quickly, and the magnetic like
force always increases, with increasing $r$. Now observe
that at the position of the sun the rotation curve is already fairly
flat \cite{Sofue:1997,Sofue:1999,Florido:2000}. Therefore on the sun the gravitational force should
be
comparable or even neglectable with respect to the magnetic-like force from the
field. Then\cite{CheungGRC0:2007}
\begin{eqnarray}
  m \frac{v^2}{r}\approx QHv
  \label{eq:eom_centripedal_magnetic_force}
\end{eqnarray}
and thus
\begin{eqnarray}
  \Omega\approx \frac{v}{r}
  \label{eq:formula_estimate_Hfield_at_sun}
\end{eqnarray}
For the sun \cite{Sofue:1997,Sofue:1999,Florido:2000}, $v_{\odot}\approx 200km\cdot s^{-1}$ and 
$r_{\odot}\approx 7.6kpc$. So
\begin{eqnarray}
  \Omega_{\odot}\approx 26s^{-1}\cdot\frac{km}{kpc}= 8.5\times10^{-16}s^{-1}
  \label{eq:result_Hfield_at_sun}
\end{eqnarray}
This is the upper limit on $\Omega$ at the Sun. Since there is still a portion
of distance further out to nearly $20kpc$ where the curve is quite flat
(with velocity $\sim200km\cdot s^{-1}$) \cite{Sofue:1997,Sofue:1999,Florido:2000}, 
we could have used these distances instead of $7.6kpc$. In that case,
it is still safe to say the upper limit of $\Omega$ is on the order 
$10^{-16}s^{-1}$. This is the strength of the field component
perpendicular to the galactic plane. To convert it to the solar system, notice 
that the north galactic pole and the north ecliptic pole form an angle of 
60.2$^\circ$ (which means the field in the solar system would be reduced almost by half
), and the milky way rotates clockwise when viewed from north galactic pole
(which means the field is negative in the solar system). Therefore from the
rotation curve of the milky way the upper limit of the effective field strength in the solar system
due to
matter in the milky way is $-10^{-17}\sim -10^{-16}s^{-1}$. This
magnitude is close to the one found by direct precession calculation
without the profile assumption, but it is not the case for field direction. The
precession indicates the field in the solar system is positive, while rotation
curve of the milky way implies it is negative. One possible explanation is that at places near the
Sun the field is dominated the positive field generated by the Sun. 
And on the other hand, the profile
fitting of precession, apart from a dipole like part due to the Sun, indeed gives a negative
background ($-0.02\times
10^{-17}s^{-1}$). However the magnitude there is smaller by almost two orders of
magnitude.

\section{$\Omega$ from rotation curves fitting of 22
galaxies}
\label{sec:22galaxy}
\begin{figure}
\begin{center}
% Use the relevant command for your figure-insertion program
% to insert the figure file.
% For example, with the option graphics use
\resizebox{0.4\textwidth}{!}{%
  \includegraphics{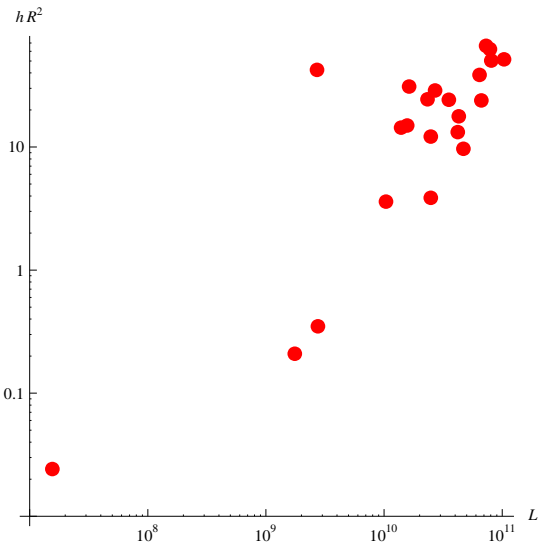}
}
% If not, use
%\vspace{5cm}       % Give the correct figure height in cm
\caption{luminosity to $\Omega R^2$ log-log plot}
\label{fig:L_hR2}       % Give a unique label
\end{center}
\end{figure}

In \cite{CheungGRC1:2008} rotation curves of 22 galaxies were fitted with the string model
and $\Omega$ is found to lie in the range $0.1\sim 16$.
There the dimension for $\Omega$ is
$s^{-1}\frac{km}{kpc}$. Thus $\Omega=1$ from fitting means 
$\Omega=3\times 10^{-17}s^{-1}$. Considering the additional factor of
$\frac{1}{2}$ in the definition of the $\Omega$ in \cite{CheungGRC1:2008}, it
means $\Omega$ using the definition of the current paper is in the range
$6\times10^{-18}\sim 1\times10^{-15}s^{-1}$ for these 22 galaxies. 
This range of magnitude covers the value in the Milky Way got from rotation curve, and also part of
those got from precesseion in the solar system. 
\paragraph{}According to the string model, considering its analogy with electromagnetism,  it is
reasonable to expect the average
field strength to be proprotional to $M^{\alpha} R^{-\beta}$ where $M$ is the
total luminous matter in the galaxy and $R$ is the size scale of the
galaxy\footnote{By the Tully-Fisher relation \cite{Tully&Fisher:1977} velocity is related to the
total
mass of the galaxy, therefore we do not need a seperate term for the velocity
dependence.}. With the assumption that $L\propto M$ where $L$ represents the
luminosity of the galaxy, it is thus interesting to see if there is any such
relation between $\Omega$, $R$ and $L$ in the fitting result for the 22 
galaxies. This serves as a consistency check for the string model. To be more
specific we will first use the electromagnetism analogy to get such a relation
theoretically as follows, and then check it with experimental data for these 22
galaxies. Consider a group of electrons azimuthal symmetrically distributed and
in rotation around the $z$ axis. Let us look at the magnetic field at the center
of this distribution\footnote{What we really want to check is the averaged field
over the galaxy, but it is proportional to the field strength at the center.}
, the determining physical quantities are: the magnetic
constant $\mu_0$, mass density scale $\rho_0$, distribution size scale $R_0$
and rotational angular velocity scale $\omega_0$. (Other determining
factors include the shape and the spatial dependence of the mass distribution, 
the spacial distribution of the angular velocity. These factors do not 
change the result of dimensional analysis but change the proportional constant.) By 
dimensional analysis, we have 
\begin{eqnarray}
  B\propto\mu_0\cdot \rho_0 \cdot\omega_0\cdot R_0^2
  \label{eq:dimensional_relation_B_and_other_factors_current_disk}
\end{eqnarray}
Using the total charge 
$Q\propto \rho_0\cdot R_0^3$, and defining $v_0=\omega_0\cdot R_0$, we have
\begin{eqnarray}
  B\propto  \mu_0\cdot Q\cdot 
  R_0^{-2}\cdot v_0
  \label{eq:dimensional_relation_B_other_factors_variables_changed}
\end{eqnarray}
Note that $v_0$ is the rotational velocity scale for the galaxy. 
Translating to the language of the string model, it is
\begin{eqnarray}
  \Omega\propto Q\cdot R_0^{-2}\cdot v_0
  \label{eq:dimensional_relation_Omega_1}
\end{eqnarray}
The proportional constant here depends only on the mass and angular velocity 
distributions, or abstractly on the galaxy type\footnote{More precisely, it is
not just the galaxy morphology type. The velocity
distribution also matters. It is possible galaxies of the same morphology type 
but very different velocity distributions will have different proportional 
constants.}. Thus galaxies with similar mass distribution profile and rotation
curve \emph{shapes} should have similar constants of proportionality. 
Now recall $M\propto Q$, where the proportional constant is universal and thus
the same for all galaxies. Furthermore we also use the
assumption\footnote{This relation is independent of the galaxy type. For more
discussion about the mass luminosity-relation among galaxies of different types,
see \cite{Roberts:1969}.} $L\propto M$. Thus
\begin{eqnarray}
  \Omega\propto L\cdot R_0^{-2} \cdot v_0
  \label{eq:dimensional_relation_OmegaR^2_Lv}
\end{eqnarray}
To relate $v_0$ to $L$ we use the Tully-Fisher relation which says $L\propto
\Delta V^{\alpha}$ where $\alpha$ is around $2.5\pm0.3$ and
$\Delta V$ is the velocity width of the galaxy\cite{Tully&Fisher:1977}. The
proportional constant in the Tully-Fisher relation is galaxy type independent. 
Since $v_0$ is the overall scale for $\Delta V$, we have also 
$L\propto (v_0)^\alpha$. Using this in (\ref{eq:dimensional_relation_OmegaR^2_Lv}),
we get
\begin{eqnarray}
  \Omega\cdot R_0^{2}\propto L^{1+\frac{1}{\alpha}}
  \label{eq:dimensional_relation_final_for_plotting}
\end{eqnarray}
which after taking logrithms
\begin{eqnarray}
  \ln\left( \Omega R_0^2 \right)=\left( 1+\frac{1}{\alpha} \right)
  \ln\left( L \right)+\ln\kappa
  \label{eq:dimensional_relation_log-log}
\end{eqnarray}
where $\kappa$ is the proportional constant in the relation
(\ref{eq:dimensional_relation_final_for_plotting}). The log-log diagram is shown in
Fig.\ref{fig:L_hR2}.

There is indeed a trend of a linear relation in the ``main'' part of the
diagram. The slope of the line is about $\frac{3}{2}$, quite close to the
$1+\frac{1}{2.5\pm0.3}$ derived theoretically. 2 points seems to lie
outside of the ``main'' part, i.e one at the lower left corner for the dwarf
galaxy m81dwb, another one for NGC4236 at the left of the upper right group. 
One possible explanation for these 2 galaxies might be that galaxy types of 
these two are too different from the rest and thus their 
$\ln\kappa$ deviated more from those in the ``main'' part. Actually in terms of
morphology type, NGC4236 is SBdm, which is the most irregular one among the
regular types, while m81dwb is the only irregular one among the 22
galaxies \cite{NED}, all others are more or less regular galaxies. Presumably 
these two lie on
another line for irregular galaxies which is parallel to the line passing the 
rest. Furthermore there might be series of parallel lines for different types
of galaxies. However, statistical error 
from the small size of this data set makes above arguments weak. On the other 
hand, it does not kill the model either. Analysis of this kind for a larger
number and more types of galaxies could make the situation clearer.

\section{Summary}
\label{sec:summary}
\paragraph{}
We analyzed the precession motion under influences of magnetic like force and
power law central force. The result for magnetic like force is later used in
computing $\Omega$ in the solar system. The result for power law central force
is tested in two examples, i.e Schwarzschild space-time and solar system with
dark matter, and found to be in agreement with analysis by other
methods. For power law central force, a simple criterian judging pro- or 
retrograde precession is derived, $n=-2$ turns out to be a critical value
between pro- and retrograde cases.
\paragraph{}
Based on the theoretical analysis of precesssion, assuming all remaining
anomalous precessions are caused by the string field, we calculated the string
field strength in the solar system\footnote{Precisely it is the upper limit of
the string field strength in the solar system.}. It is found to be on the order
of $10^{-17}s^{-1}$. Except the data of the Saturn, anomalous precessions of
other planets considered in this paper indicate a string field of almost 
$\frac{1}{r^3}$ decreasing pattern, where $r$ is the distance to the Sun. 
\paragraph{}
For the string field strength in the solar system, a profile
fitting is further done to compare with the result directly calcualted from
precession. We assumed the string field consists of two parts. One from the Sun,
which goes as $\frac{1}{r^a}$. The second part comes from the rest of the
Milky Way, which is a constant background within the solar system region.
The part due to the Sun is found to be nearly a dipole field, i.e
$\frac{1}{r^3}$. The background is found to be negative, which, when combined
with analysis using electromagnetism analogy, correctly matches with the
fact that the solar system and the Milky Way rotate in opposite directions.
\paragraph{}
As an independent check of the calculated result of the string field in the 
solar system, we used the rotation curve of the Milky Way to estimate the string
field strength in the Milky Way, which contains the solar system. The estimation
gives a similar order of magnitude with that got from precessions. But the direction is opposite. If
we consider the result from profile fitting, then the background direction agrees with the one
predicted by Milky Way rotation curve, but the magnitude is smaller by 2 orders of magnitude.
\paragraph{}
For 22 galaxies fitted in \cite{CheungGRC1:2008}, the string field strength is found to span
a range just around that of the solar system, or the Milky Way. Note that here the
strength is found by precise fitting of rotation curves of the 22 galaxies, which
is a completely independent method from fitting precession.
\paragraph{}
Based on an analysis of an electromagentism analogy, we derived a relation
between $\Omega R^2$ and luminosity. This relation is then checked by fitting
result from \cite{CheungGRC1:2008}. Except for two galaxies, the data for 
the other 20 galaxies all satisfy the derived relation to a good degree. This way the string model
survived one model consistency check.
\paragraph{}
In summary of analysis about the string field strength $\Omega$, we used three 
kinds of observations to find corresponding values of
$\Omega$: 1) rotation curves of 22 galaxies, 2) rotation curve of the Milkyway
and 3) anomalous precessions in the solar system(, which is also in the 
Milkyway). All the situations are summarized in Table.\ref{tab:paper_summary},
and we can see there is no obvious inconsistency in using the string model for 
various objects and obsevations.
\begin{table*}
\begin{center}
  \begin{tabular}{c||cc|c}
\hline
    & Solar System $\subset$ & Milky Way 
    & 22 other galaxies\\\hline
    Rotation Curve & & $|\Omega|\lesssim10^{-16}$ 
    &$6\times10^{-18}\lesssim|\Omega|\lesssim10^{-15}$\\
    Precession &$|\Omega|\lesssim 10^{-17}$ & &\\\hline
  \end{tabular}
  \caption{\label{tab:paper_summary}Summary (Dimension for $\Omega$ is $s^{-1}$). The first
 column indicates the kind of observation used.}  
\end{center}
\end{table*}

\section*{Acknowledgement}
\label{sec:acknowledgement}
The author would like to thank Yuran Chen and Youhua Xu for colaboration at the
early stage of this work. The author would also like to thank Edna Cheung for 
suggesting study the model by precession, try the profile fitting and a 
discussion that triggered the effort to derive the relation between $\Omega R^2$
and $L$. The author has also benefited from discussions with Anke Knauf, Lingfei
Wang, Tianheng Wang and Yun Zhang. This work is supported by the National Science Foundation of
China under the Grant No. 0204131361.

%\appendix
\section*{Appendix A: Solar system magnetic field and its effect on planet perihelion precession}
\label{appenix:A}
References for this appendix are
\cite{Parker,Parker:1958,Encyclopedia:1997,Meyer:Parker:Simpson:1956,Babcock:1961}. The Sun and most
planets in the solar system have magnetic field due to dynamo
effect. If we treat both the Sun and the planet as magnetic dipoles interacting
in vaccum (which leads to a central force with $n=-4$), using data of magnetic
fields of the Sun (around $1\sim 2$ gauss at the polar region) and the Earth 
(around 0.6 gauss at the polar region), we can find the corresponding precession
produced is nearly $10^{-6}$arcsec/cy, which is 2 orders of magnitude smaller than
the observed one. In fact the magnetic field in the solar system is much more
complicated than those produced by several dipoles in vacuum. First the solar
system is not empty but filled with particles emitted from the Sun, i.e the solar
wind. Charged particles lock with it the magnetic field of the Sun and spread it
all around in the solar system. From the Sun to about the position of the Earth,
the magnetic force line is parrell to the radial stream of solar wind particle 
and falls off by $r^{-2}$. From the position of the Earth to about
position of the Mars is a field free region with $B<10^{-6}$ gauss. Further out
to the position of the Jupiter is a region with disordered magnetic field with
$B\sim 10^{-5}$ gauss. For precession, the most important feature of the solar
system field is that it is oscillating. Firstly, for the Sun the magnetic dipole
axis is inclined relative to the rotational axis. This leads to an oscillating
neutral current sheet. Therefore planets on the ecliptic plane is above the
netral current sheet for half of solar self rotation period, below for another half.
Since field directions above and below the netral current sheet is opposite, the
magnetic force experienced by the planets also change directions within one self
rotation of the Sun. Secondly, the magnetic field of the Sun also changes
direction every 22 years due to its differential rotation, which leads to another
oscillation of magnetic field on the planets. Alltogether these two
oscillations render the magnetic field effect on precession neglectable with 
respect to other accumulating effects. 

% BibTeX users please use
 \bibliographystyle{epj.bst}
 \bibliography{reference}

\begin{thebibliography}{19}
\providecommand{\natexlab}[1]{#1}
\providecommand{\url}[1]{\texttt{#1}}
\expandafter\ifx\csname urlstyle\endcsname\relax
  \providecommand{\doi}[1]{doi: #1}\else
  \providecommand{\doi}{doi: \begingroup \urlstyle{rm}\Url}\fi

\bibitem[Cheung et~al.(2007)Cheung, Savvidy, and Kao]{CheungGRC0:2007}
Yeuk-Kwan~E. Cheung, Konstantin Savvidy, and Hsien-Chung Kao.
\newblock Galaxy rotation curves from string theory.
\newblock Feb 2007.

\bibitem[Cheung and Xu(2008)]{CheungGRC1:2008}
Yeuk-Kwan~E Cheung and Feng Xu.
\newblock Fitting the galaxy rotation curves: Strings versus nfw profile.
\newblock Oct 2008.

\bibitem[Iorio(2009)]{Iorio:2009}
Lorenzo Iorio.
\newblock \emph{The Astronomical Journal 137 3615}, 2009.

\bibitem[Goldstein et~al.(2002)Goldstein, Poole, and
  Safko]{Goldstein:mechanics}
Herbert Goldstein, Charles~P. Poole, and John~L. Safko.
\newblock \emph{Classical Mechanics, 3rd Edition}.
\newblock Addison Wesley, 2002.

\bibitem[Weinberg(1972)]{Weinberg:gravitation}
Steven Weinberg.
\newblock \emph{{Gravitation and cosmology: principles and applications of the
  general theory of relativity}}.
\newblock Wiley, 1972.

\bibitem[Misner et~al.(1973)Misner, Thorne, and Wheeler]{Misner:gravitation}
CW~Misner, KS~Thorne, and JA~Wheeler.
\newblock \emph{Gravitation}.
\newblock W. H. Freeman, 1973.

\bibitem[Gr{\o}n and Soleng(1995)]{Gron:1995}
{\O}yvind Gr{\o}n and Harald~H. Soleng.
\newblock Experimental limits to the density of dark matter in the solar
  system.
\newblock 1995.

\bibitem[HOR()]{HORIZON}
Horizon web interface.
\newblock URL \url{http://ssd.jpl.nasa.gov/?horizons}.

\bibitem[Parker(1958)]{Parker}
E.~N. Parker.
\newblock dynamics of the interplanetary gas and magnetic fields.
\newblock \emph{Astrophysical Journal}, 128, 1958.

\bibitem[Honma and Sofue(1997)]{Sofue:1997}
Mareki Honma and Yoshiaki Sofue.
\newblock Rotation curve of the galaxy.
\newblock \emph{Publ. Astron. Soc. Japan}, 49, 1997.

\bibitem[Sofue et~al.(1999)Sofue, Tutui, Honma, Tomita, Takamiya, Koda, and
  Takeda]{Sofue:1999}
Y.~Sofue, Y.~Tutui, M.~Honma, A.~Tomita, T.~Takamiya, J.~Koda, and Y.~Takeda.
\newblock Central rotation curves of spiral galaxies.
\newblock 1999.

\bibitem[Battaner and Florido(2000)]{Florido:2000}
E~Battaner and E~Florido.
\newblock The rotation curve of spiral galaxies and its cosmological
  implications.
\newblock \emph{Fund. Cosmo Phys.}, 21, 2000.

\bibitem[Tully and Fisher(1977)]{Tully&Fisher:1977}
R.~B. Tully and J.~R. Fisher.
\newblock A new method of determining distances to galaxies.
\newblock \emph{J. R. Astronomy and Astrophysics, vol. 54, no. 3}, pages
  661--673, Feb 1977.

\bibitem[Roberts(1969)]{Roberts:1969}
M.S. Roberts.
\newblock \emph{Astronomical Journal}, 74, 1969.

\bibitem[NED()]{NED}
Nasa/ipac extragalactic database.
\newblock URL \url{http://nedwww.ipac.caltech.edu}.

\bibitem[E.N.Parker(1958)]{Parker:1958}
E.N.Parker.
\newblock Cosmic-ray modulation by solar wind.
\newblock \emph{Physical Review}, 110, 1958.

\bibitem[Shirley and Fairbridge(1997)]{Encyclopedia:1997}
J.~H. Shirley and R.~W. Fairbridge.
\newblock \emph{Encyclopedia of Planetary Sciences}.
\newblock September 1997.

\bibitem[Meyer et~al.(1956)Meyer, Parker, and
  Simpson]{Meyer:Parker:Simpson:1956}
P.~Meyer, E.~N. Parker, and J.~A. Simpson.
\newblock Solar cosmic rays of february, 1956 and their propagation through
  interplanetary space.
\newblock \emph{Physical Review}, 104, 1956.

\bibitem[Babcock(1961)]{Babcock:1961}
H.~W. Babcock.
\newblock The topology of the sun's magnetic field and the 22-year cycle.
\newblock \emph{Astrophysical Journal}, 133, 1961.

\end{thebibliography}

\end{document}